\documentclass[sigconf,10pt]{acmart}

\usepackage[utf8]{inputenc} %

\usepackage{graphicx}

\usepackage[T1]{fontenc}

\IfFileExists{headers/config/showoverfull.config}{
	\overfullrule=1cm
}{
}

\usepackage{etoolbox}

\newbool{includeappendix}
\setbool{includeappendix}{true} %
\IfFileExists{headers/config/noappendix.config}{
	\setbool{includeappendix}{false}
}{}

\usepackage{xr} %
\usepackage{filecontents}

\ifbool{includeappendix}{}{

	\externaldocument{appendix-labels}
}

\newcommand{\eg}{e.g.,\xspace}
\newcommand{\ie}{i.e.,\xspace}
\newcommand{\etc}{etc.\xspace}

\usepackage{caption}
\usepackage{subcaption}

\usepackage{xcolor} %

\definecolor{my-full-blue}{HTML}{1F77B4}

\definecolor{my-full-orange}{HTML}{FF7F0E}

\definecolor{my-full-green}{HTML}{2CA02C}

\definecolor{my-full-red}{HTML}{d62728}

\definecolor{my-full-purple}{HTML}{9467bd}

\definecolor{my-full-brown}{HTML}{8c564b}

\definecolor{my-full-pink}{HTML}{e377c2}

\definecolor{my-full-gray}{HTML}{7f7f7f}

\definecolor{my-full-olive}{HTML}{bcbd22}

\definecolor{my-full-cyan}{HTML}{17becf}

\colorlet{my-blue}{my-full-blue!30}
\colorlet{my-orange}{my-full-orange!30}
\colorlet{my-green}{my-full-green!30}
\colorlet{my-red}{my-full-red!30}
\colorlet{my-purple}{my-full-purple!30}
\colorlet{my-brown}{my-full-brown!30}
\colorlet{my-pink}{my-full-pink!30}
\colorlet{my-gray}{my-full-gray!30}
\colorlet{my-olive}{my-full-olive!30}
\colorlet{my-cyan}{my-full-cyan!30}

\definecolor{w1color}{rgb}{0.2980392156862745, 0.4470588235294118, 0.6901960784313725}
\definecolor{w2color}{rgb}{0.8666666666666667, 0.5176470588235295, 0.3215686274509804}
\definecolor{w3color}{rgb}{0.3333333333333333, 0.6588235294117647, 0.40784313725490196}

\definecolor{pastelblue}{rgb}{0.6588235294117647, 0.8392156862745098, 1}

\spaceskip=1\fontdimen2\font plus 1\fontdimen3\font minus 2\fontdimen4{\bfseries}{\fontseries{b}\selectfont}

\usepackage[english]{babel}
\usepackage{balance}

\usepackage{amsmath}
\DeclareMathAlphabet{\mathcal}{OMS}{cmsy}{m}{n}  %
\usepackage{booktabs}
\usepackage{xspace}
\usepackage{enumitem}

\usepackage{siunitx}

\definecolor{lightblue}{RGB}{212,234,248}
\definecolor{darkblue}{RGB}{29,117,180}
\definecolor{lightgreen}{RGB}{204,255,204}
\definecolor{darkgreen}{RGB}{101,178,50}
\definecolor{orange}{RGB}{255,171,89}

\usepackage[most]{tcolorbox}

\usepackage{tabularx}
\newcommand{\smallindent}{\hphantom{N}}

\DeclareSIUnit{\nothing}{\relax}

\newcommand{\first}{\emph{(i)}\xspace}
\newcommand{\second}{\emph{(ii)}\xspace}
\newcommand{\third}{\emph{(iii)}\xspace}

\newcommand*{\researchquestion}[1]{%
	\begin{tcolorbox}[%
			boxsep=4pt, left=0pt,right=0pt,top=0pt,bottom=0pt,
			width=\columnwidth,
			colframe=pastelblue!50, colback=pastelblue!50,
			arc=0.5mm,
		]
		\centering
		#1%
	\end{tcolorbox}
}

\makeatletter
\renewcommand\paragraph{\@startsection{paragraph}{4}{\z@}%
	{-.5\baselineskip \@plus -2\p@ \@minus -.2\p@}%
	{-3.5\p@}%
	{\bfseries\@addspaceafter}}
\makeatother

\usepackage[capitalize]{cleveref}

\makeatletter
\newcommand\theHALG@line{\thealgorithm.\arabic{ALG@line}}
\makeatother

\crefformat{section}{\S#2#1#3}

\crefrangeformat{section}{\S#3#1#4\crefrangeconjunction\S#5#2#6}

\crefmultiformat{section}{\S#2#1#3}{\crefpairconjunction\S#2#1#3}{\crefmiddleconjunction\S#2#1#3}{\creflastconjunction\S#2#1#3}

\newcommand{\crefrangeconjunction}{--}

\crefname{listing}{Lst.}{listings}
\crefname{line}{Lin.}{Lin.}
\crefname{appendix}{Appendix}{Appendix}

\newcommand{\appref}[1]{%
	\ifbool{includeappendix}{\cref{#1}}{the appendix}%
}
\newcommand{\Appref}[1]{%
	\ifbool{includeappendix}{\cref{#1}}{The appendix}%
}

\usepackage[normalem]{ulem}

\title{A New Hope for Network Model Generalization}

\author{\mbox{Alexander Dietmüller}{$^*$}}  %
\orcid{0000-0003-3769-3958}
\affiliation{%
ETH Zürich
}

\author{Siddhant Ray}
\orcid{0000-0003-0265-2144}
\affiliation{%
ETH Zürich
}

\author{Romain Jacob}
\orcid{0000-0002-2218-5750}
\affiliation{%
ETH Zürich
}

\author{Laurent Vanbever}
\orcid{0000-0003-1455-4381}
\affiliation{%
ETH Zürich
}

\copyrightyear{2022}
\acmYear{2022}
\setcopyright{rightsretained}
\acmConference[HotNets '22]{The 21st ACM Workshop on Hot Topics in Networks}{November 14--15, 2022}{Austin, TX, USA}
\acmBooktitle{The 21st ACM Workshop on Hot Topics in Networks (HotNets '22), November 14--15, 2022, Austin, TX, USA}
\acmDOI{10.1145/3563766.3564104}
\acmISBN{978-1-4503-9899-2/22/11}

\begin{document}

\begin{abstract}
    Generalizing machine learning (ML) models for network traffic dynamics tends to be considered a lost cause.
    Hence for every new task, we design new models and train them on model-specific datasets closely mimicking the deployment environments.
    Yet, an ML architecture called \emph{Transformer} has enabled previously unimaginable generalization in other domains.
    Nowadays, one can download a model pre-trained on massive datasets and only fine-tune it for a specific task and context with comparatively little time and data. These fine-tuned models are now state-of-the-art for many benchmarks.

    We believe this progress could translate to networking and propose a Network Traffic Transformer (NTT), a transformer adapted to learn network dynamics from packet traces.
    Our initial results are promising: NTT seems able to generalize to new prediction tasks and environments.
    This study suggests there is still hope for generalization
    through future research.
\end{abstract}



\maketitle
\renewcommand{\shortauthors}{A.\ Dietmüller, S.\ Ray, R.\ Jacob, and L.\ Vanbever}

\vspace{-3mm}
{\small {$^*$}The \href{https://credit.niso.org/}{CRediT statement} for this work is available at~\cite{CRediT}.}

\section{Introduction}
\label{sec:intro}

Modeling network dynamics is a \emph{sequence modeling} problem: From a sequence of past packets, estimate the current state of the network (\eg Is it congested?), then predict the state's evolution and future traffic's fate---or which action to take next.
This is a notoriously complex task, and the research community is increasingly turning to Machine Learning (ML) for solutions in many applications, including
congestion control~\cite{abbaslooClassicMeetsModern2020,jayDeepReinforcementLearning2019,nieDynamicTCPInitial2019,winsteinTCPExMachina2013},
video streaming~\cite{akhtarOboeAutotuningVideo2018,maoNeuralAdaptiveVideo2017,yanLearningSituRandomized2020},
traffic optimization~\cite{chenAuTOScalingDeep2018},
routing~\cite{valadarskyLearningRoute2017},
flow size prediction~\cite{dukicAdvanceKnowledgeFlow2019,poupartOnlineFlowSize2016},
MAC protocol optimization~\cite{jogOneProtocolRule2021,yuDeepReinforcementLearningMultiple2019},
and network simulation~\cite{zhangMimicNetFastPerformance2021}.

\begin{figure}
    \centering
    \includegraphics[scale=1]{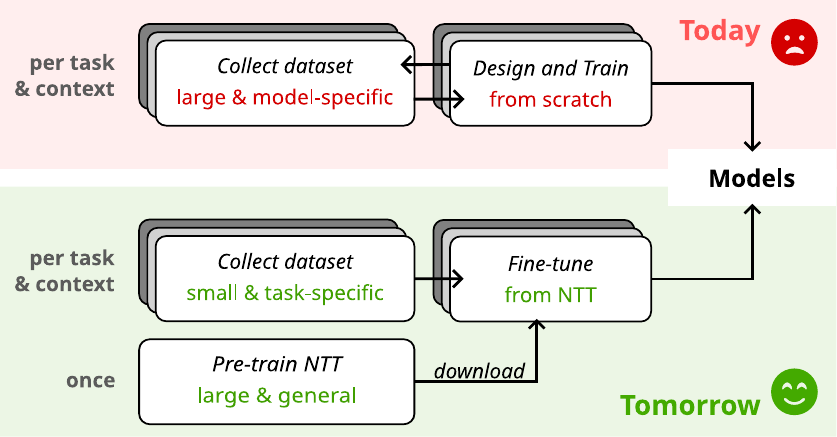}
    \caption{Could we collectively learn general network traffic dynamics \emph{once} and focus on task-specific data collecting and learning for \emph{all future models?}}
    \label{fig:vision}
\end{figure}

\paragraph{Problem}
Today's models do not generalize well;
\ie they often fail to deliver outside of their original training environments~\cite{yanLearningSituRandomized2020,bakshyRealworldVideoAdaptation2019,bartulovicBiasesDataDrivenNetworking2017,yanPantheonTrainingGround2018,fuUseMLBlackbox2021}; generalizing to different tasks is not even considered.
Recent work argues that, rather than hoping for generalization, one obtains better results by training in-situ, \ie using data collected in the deployment environment~\cite{yanLearningSituRandomized2020}.
Thus, today we tend to design and train models from scratch using model-specific datasets~(\cref{fig:vision}, top).
This process is repetitive, expensive, and time-consuming.
Moreover, the growing resource requirements to even attempt training these models is increasing inequalities in networking research and, ultimately, hindering collective progress.

\paragraph{Vision}
We argue there is still hope for generalization in networking. Even if the networking contexts (topology, network configuration, traffic, \etc) are very diverse, the underlying dynamics remain similar; \eg when buffers fill up, queuing disciplines delay or drop packets.
These dynamics can be learned with ML, and there is no need to relearn everything every time, \eg how congestion looks like.
\begin{quote}
    We envision a \emph{generic network model} trained to capture the shared dynamics underpinning any network---once---which can be fine-tuned for many different networking tasks and contexts.
\end{quote}
Tackling this challenge would benefit the entire community.
Starting from such a model, one would only need to collect a small task-specific dataset for fine-tuning~(\cref{fig:vision}, bottom), assuming that the pre-trained model generalizes well.

Existing approaches---while not performing \emph{optimally} outside of their training environments---provide evidence for generalization.
The congestion control algorithm Aurora~\cite{jayDeepReinforcementLearning2019} performs adequately in environments with bandwidth more than an order of magnitude higher than during training.
Models trained with Genet~\cite{xiaAutomaticCurriculumGeneration2022} on simulation data perform well in several real-world settings.
But truly ``generic'' models---able to perform well on a wide range of tasks and networks---remain unavailable, as mixing different contexts is unpredictable.
In some cases, more diverse training data has been shown to provide benefits without consequences, \eg training over a wide range of propagation delays in~\cite{sivaramanExperimentalStudyLearnability2014}.
Yet in other cases, mixing contexts can decrease performance,\eg varying numbers of senders in~\cite{sivaramanExperimentalStudyLearnability2014} or wired and wireless traces in~\cite{bartulovicBiasesDataDrivenNetworking2017},
if the model is not able to tell these contexts apart.

\paragraph{Game-changer}
A few years ago, a new architecture for sequence modeling was proposed: the \emph{Transformer}~\cite{vaswaniAttentionAllYou2017}.
This architecture is designed to train efficiently,%
\footnote{%
    {Transformers scale better than recurrent neural networks, another popular architecture for sequence modeling that Transformers effectively succeeded.}
} enabling learning from massive datasets and unprecedented generalization across \emph{multiple} contexts.
In a pre-training phase, the transformer learns contextual sequential ``structures,'' \eg the structure of a language from a large corpus of texts.
Then, in a much quicker fine-tuning phase, the final stages of the model are adapted to a specific prediction task.
Today, transformers are among the state-of-the-art in natural language processing (NLP~\cite{storksRecentAdvancesNatural2020}) and computer vision (CV~\cite{hanSurveyVisionTransformer2022,khanTransformersVisionSurvey2021}).

Transformers generalize well because they can learn to distinguish different contexts during pre-training;
they learn rich contextual representations~\cite{devlinBERTPretrainingDeep2019} where the representation of the same element, \eg a word, depends on its context, inferred from the sequence.
Consider two input sequences:
\emph{Stick to it!} and \emph{Can you hand me this stick?}
The transformer output for each \emph{stick} is different as it encodes the word's context.
This contextual output is an efficient starting point for fine-tuning the model to diverse downstream tasks, \eg question answering, text comprehension, or sentence completion~\cite{storksRecentAdvancesNatural2020}.
We can draw parallels between networking and NLP: packet metadata alone (headers, delay, \etc) provide limited insights into the network state---we also need the context, \ie the history of past packets.
For example, increasing latency indicates congestion, and loss patterns or ACK batching may allow for differentiating wired and wireless connections.

We believe a transformer can learn \emph{many} such network-specific contexts.
If it does, it could pave the way for generalization in networking, as it did for NLP and CV.

\paragraph{Challenges}
Naively transposing NLP or CV transformers to networking fails, unsurprisingly. We must adapt them to the peculiarity of networks.
In particular, ``sequences'' must be carefully defined: While text snippets and images are relatively self-contained, any packet trace only gives a partial view of the network.
Moreover, generalizing the diversity and dynamism of protocol interactions is far from trivial.
Ultimately, we identify three main open questions.
\begin{enumerate}[noitemsep]
    \item
          How to adapt transformers for networking?
    \item
          Which pre-training task would allow the model to generalize, and how far can we push generalization?
    \item
          How to assemble a dataset large and diverse enough to allow useful generalization?
\end{enumerate}

\vspace*{-1mm} %
\paragraph{Contributions}
After a short background on transformers (\cref{sec:background}), we begin to answer these questions and present NTT: our proof-of-concept \emph{Network Traffic Transformer}~(\cref{sec:transformer}, \cite{NTTsoftware}).
Preliminary simulations~(\cref{sec:eval}) provide first evidence that NTT can learn network traffic dynamics and generalize to new tasks and environments.
This opens a broad research agenda~(\cref{sec:future}).

\section{Background on Transformers}
\label{sec:background}

In this section, we introduce \emph{attention}, the mechanism behind Transformers; detail the idea of pre-training and fine-tuning; and present insights from adapting Transformers to CV.

\paragraph{Sequence modeling with attention}
Transformers are built around the \emph{attention} mechanism, which maps an input sequence to an output sequence of the same length.
Every output encodes its own information \emph{and its context}, \ie information {inferred} from related elements in the sequence.
For a detailed explanation, we refer to~\cite{vaswaniAttentionAllYou2017} and excellent online guides~\cite{huangAnnotatedTransformer,alammarIllustratedTransformer}.
Computing attention is efficient as all elements in the sequence can be processed in parallel with matrix operations that are highly optimized on most hardware.

\begin{figure}[t]
    \hfill
    \begin{subfigure}{0.43\columnwidth}
        \includegraphics[scale=1]{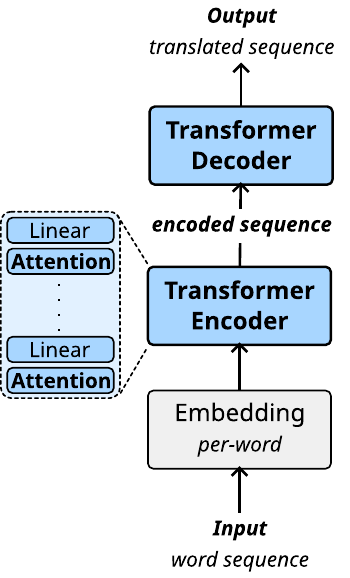}
        \caption{Transformer\\}
        \label{fig:transformers_original}
    \end{subfigure}\hfill%
    \begin{subfigure}{0.25\columnwidth}
        \includegraphics[scale=1]{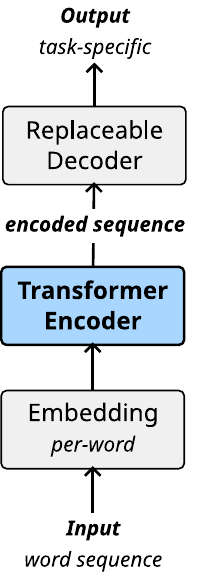}
        \caption{BERT}
        \label{fig:transformers_bert_pre}
    \end{subfigure}\hfill%
    \begin{subfigure}{0.25\columnwidth}
        \includegraphics[scale=1]{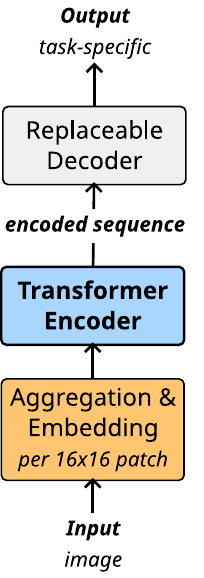}
        \caption{ViT}
        \label{fig:transformers_vit}
    \end{subfigure}\hfill
    \captionsetup{skip=\dimexpr\abovecaptionskip-2mm}
    \caption{Transformer variants.}
\end{figure}

While attention originated as an improvement to recurrent neural networks (RNNs), Vaswani et al.~\cite{vaswaniAttentionAllYou2017} realized that it could replace them entirely.
The authors propose an architecture for translation tasks that contains:
an \emph{embedding} layer mapping words to vectors;
a \emph{transformer encoder} encoding the input sequence;
and a \emph{transformer decoder} generating an output sequence based on the encoded input~(\cref{fig:transformers_original}).
Each transformer block alternates between attention and linear layers, \ie between encoding context and refining features.

\paragraph{Pre-training and fine-tuning}
Transformers are used for a wide range of NLP tasks, and the prevailing strategy is to use pre-training and fine-tuning.
We explain this approach on the example of BERT~\cite{devlinBERTPretrainingDeep2019}, one of the most widely used transformer models.
BERT uses only the transformer encoder, followed by a small and replaceable decoder.%
\footnote{%
    Usually, a multilayer perceptron (MLP) with a few linear layers; this decoder is often called the `MLP head'.
}
BERT is \emph{pre-trained} with a task that requires learning language structure.
Concretely, a fraction of words in the input sequence is masked out, and the decoder is tasked to predict the original words from the encoded input sequence~(\cref{fig:transformers_bert_pre}).
Conceptually, this is only possible if the encoding includes sufficient \emph{context} to infer the missing word.
Afterward, the unique pre-trained model can be fine-tuned to many different tasks by replacing the small decoder with task-specific ones, \eg language understanding, question answering, or text generation~\cite{devlinBERTPretrainingDeep2019,zaibBERTCoQACBERTbasedConversational2021, chenDistillingKnowledgeLearned2020}.
The model has already learned to encode language context and only needs to learn to extract the task-relevant information from this context.
This requires far less data compared to starting from scratch:
BERT is pre-trained from text corpora with several billion words and fine-tuned with $\sim$100 thousand examples per task.
Furthermore, BERTs pre-training task is unsupervised, \ie it requires only ``cheap'' unlabeled data for masking and reconstruction.
``Expensive'' labeled data, \eg for text classification, is only needed for fine-tuning.

\paragraph{Vision transformers}
Following their success in NLP, Transformers gained traction in CV as well, with two notable distinctions:
\first input aggregation;
and \second a domain-specific pre-training task.
While attention is efficient to parallelize, it needs to compare each element in the sequence with each other element to encode context.
Consequently, the required computation scales quadratically with the input sequence length, and using sequences of individual pixels does not scale to images of high resolution.
As a solution, the Vision Transformer~(ViT, \cite{dosovitskiyImageWorth16x162021,beyerBetterPlainViT2022}) aggregates pixels into 16$\times$16 patches and applies the embedding and transformer layers to the resulting sequence of patches, using an architecture similar to BERT~(\cref{fig:transformers_vit}).
However, using a classification task to pre-train ViT delivered better results than a reconstruction task.
This shows the importance of domain-appropriate pre-training:
it may be possible to reconstruct a patch by only considering neighboring ones, but classification requires understanding the whole image, \ie the context of the entire sequence.

\section{Network Traffic Transformer}
\label{sec:transformer}

\begin{figure}[b]
    \centering
    \includegraphics[scale=1]{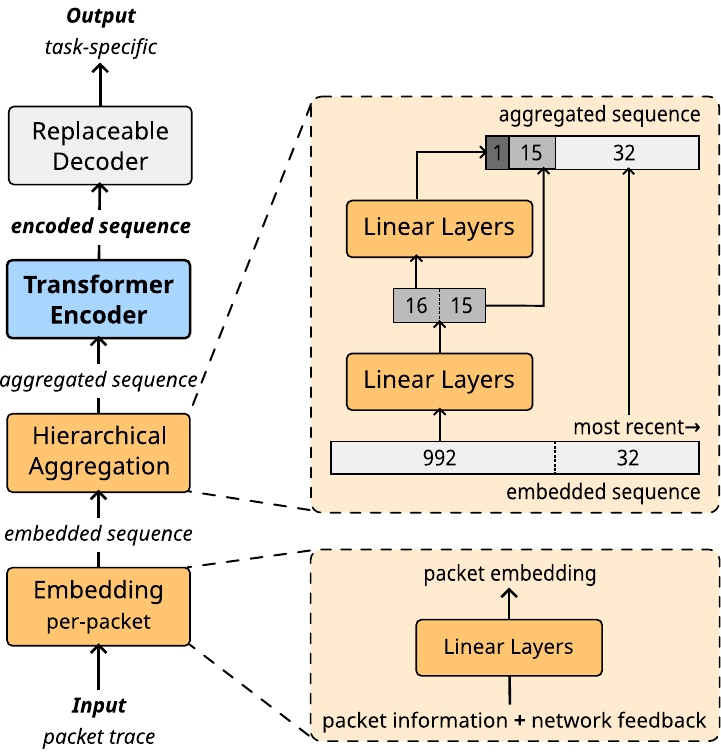}
    \caption{%
        The Network Traffic Transformer (NTT) contains three main stages:  %
        embedding,
        aggregation, and
        a transformer encoder.
        It outputs a context-rich encoded sequence that is fed into a task-specific decoder.
    }
    \label{fig:architecture_overview}
\end{figure}

Given the success of Transformers in NLP and CV and the similarities between the underlying sequence modeling problems, we postulate that transformers could also generalize network traffic dynamics.
This section presents our proof-of-concept: the Network Traffic Transformer (NTT, \cref{fig:architecture_overview}).

\begin{enumerate}[noitemsep]
    \item
          Packets are more complex than words or pixels.
          Which packet features are helpful and which are necessary to learn network dynamics?
    \item
          The fate of a packet may depend on much older ones.
          As the sequence length is practically limited~(\cref{sec:background}), how can we capture both short- and long-term network dynamics in the input sequence?
    \item
          Which pre-training task enables the model to learn general network patterns effectively?
\end{enumerate}

\paragraph{Learning feature extraction}

Packets carry a lot of information that could be used as model features, \eg header fields.
Today, we typically use domain knowledge to manually extract and aggregate features and feed these into off-the-shelf ML architectures.
We argue this is sub-optimal for two reasons:
\first
we select features for a specific task and dataset, which limits generalization;
\second
since the features are not learned from data, they may end up sub-optimal for the task.

Instead, we propose to let the model learn useful features from raw data. To learn traffic dynamics from a sequence of packets, we must provide the model with information about the packets as well as their fate in the network.
Since we do not want to define a priori how important the individual pieces of information are, we feed them all into a first embedding layer~(\cref{fig:architecture_overview}).
It is applied to every packet separately.

In our proof-of-concept, we use minimal information:
\emph{timestamp},
\emph{packet size},
\emph{receiver ID},%
\footnote{%
    An IP addresses proxy, as we do not want to learn IP address parsing (yet).
}
and \emph{end-to-end delay}.
These enable learning embeddings with temporal (delays over time) and spatial (impact of packet size on delay) patterns.
We discuss the challenge of embedding more information in~\cref{sec:future}.

\paragraph{Learning packet aggregation}

Packet sequences must be sufficiently long to capture more than short-term traffic dynamics.
But as the training time of Transformers scales quadratically with the sequence length, we face practical limitations.

We address this problem by using a hierarchical aggregation layer~(\cref{fig:architecture_overview}).
We aggregate a long packet sequence into a shorter one while letting the model learn how to aggregate the relevant historical information, similar to the pixel patch aggregation in ViT~\cite{dosovitskiyImageWorth16x162021}.
However, we aim to both aggregate \emph{and} retain recent packet-level details.
To achieve this, we keep the most recent packets without aggregation and the longer traffic is in the past, the more we aggregate, as details become less relevant to predict the current traffic dynamics.

In our proof-of-concept, we set the input sequence length to 1024 packets, enough to cover the number of in-flight packets in our experiments.
We aggregate this sequence into 48 elements in two stages:
the most recent packets are kept as-is; less recent packets are aggregated once; and the least recent twice~(\cref{fig:architecture_overview}).
Our multi-timescale aggregation is easy to adapt to a larger history without sacrificing recent packet details.
We show in \cref{sec:eval} that this aggregation is beneficial, but it is unclear which sequence length and levels of aggregation generalize best; we discuss this further in~\cref{sec:future}.

\paragraph{Learning network patterns}

Finally, we need a training task that allows NTT to learn network dynamics: in our proof-of-concept, we use end-to-end delay prediction.
We aim to pre-train NTT to generalize to a large set of fine-tuning tasks.
Consequently, we need a pre-training task that is generic enough to be affected by many network effects.
As almost everything in a network affects packet delays (\eg path length, buffer sizes), a delay prediction task seems a rational choice.

To pre-train NTT, we mask the delay of the most recent packet in the sequence and use a decoder with linear layers to predict the actual delay.
During training, the NTT must learn which features are useful (embedding layer), how to aggregate them over time (aggregation layer), and
to infer context from the whole sequence (transformer encoder layers).

During fine-tuning, one can update or replace the decoder~(\cref{fig:architecture_overview}) to adapt NTT to a new environment (\eg same decoder in a different network) or to new tasks (\eg predicting message completion times).
This is efficient as the knowledge accumulated by NTT during pre-training generalizes well to the new task, as we demonstrate in the next section.

\section{Preliminary Evaluation}
\label{sec:eval}

Our preliminary evaluation of NTT in simulation shows that:
\begin{enumerate}[noitemsep]
    \item NTT is able to learn some network dynamics;
    \item Pre-training helps to generalize;
    \item Networking-specific design helps generalization.
\end{enumerate}

\noindent
Importantly, we do \emph{not} aim to show that NTT outperforms existing specialized models (yet%
\footnote{%
    The research on Transformers in CV showed that large datasets are required for transformers to outperform the state-of-the-art.
}%
).
Here, we focus on assessing the potential of our approach.

\paragraph{Datasets}

\begin{figure}[t]
    \centering
    \includegraphics[scale=1]{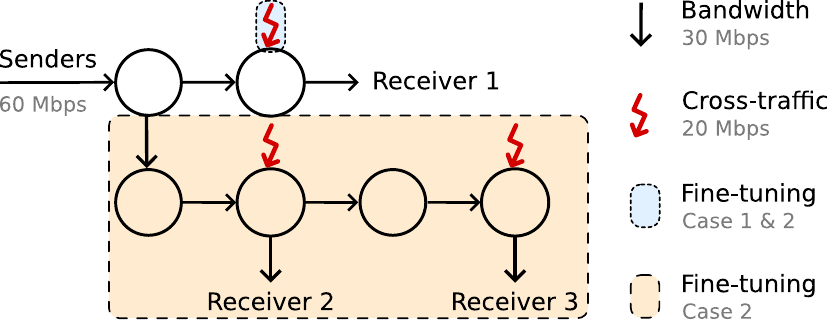}
    \caption{Dataset generation setup.}
    \label{fig:ns3}
\end{figure}

We use ns-3~\cite{rileyNs3NetworkSimulator2010} and the setup in \cref{fig:ns3} to generate several datasets; one for pre-training, and several for fine-tuning. From each, we reserve a fraction for testing.

In the pre-training dataset, 60 senders generate 1Mbps of messages each,  following real-world traffic distributions~\cite{montazeriHomaReceiverdrivenLowlatency2018}.
They send messages over a bottleneck link with 30Mbps bandwidth and a queue size of 1000 packets.
We run 10 simulations for 1 minute each with randomized application start times.
This dataset contains about 1.2 million packets.

For the fine-tuning datasets, we add cross-traffic (case 1) and additionally extend the network topology (case 2). Cross-traffic is modeled as 20Mbps of TCP flows.
Note that the datasets do \emph{not} contain the cross-traffic packets, \emph{only} those from the senders.
For each case, we generate a dataset containing roughly as many packets as the pre-training dataset and a ``smaller'' dataset containing about 10\% of the packets.

\paragraph{Models}
We compare several versions of NTT.
The \emph{pre-trained} models first learn from the pre-training and then one fine-tuning dataset, while the \emph{from scratch} versions only learn from one fine-tuning dataset.
We also pre-train ablated versions of NTT: we compare our multi-timescale aggregation of 1024 packets into 48 aggregates (see \cref{sec:transformer}) with
\emph{no aggregation} (using only 48 individual packets) and
\emph{fixed aggregation} (using 48 aggregates of 21 packets each, \ie 1008-packet sequences).
In addition, we pre-train one model \emph{without delay} and one \emph{without packet size} information in the input sequences.
Finally, we consider two naive baselines:
one always returns the \emph{last observed} output value; another returns
an \emph{EWMA}.%
\footnote{%
    Exponentially Weighted Moving Average; we used $\alpha=0.01$.
}

\paragraph{Tasks}
We evaluate our models on two prediction tasks.
The first is to predict the delay of the last packet of the sequence;
this task is also used for pre-training.
The second task is to predict the message completion times (MCTs), \ie the time until the final packet of a message is delivered.
This flow-level prediction task uses a decoder with two inputs:
the NTT outputs for the past packets and the message size.
We report the mean-squared error (MSE) for both tasks and process MCTs on a logarithmic scale to limit the impact of outliers.%
\footnote{%
    MCT mean: 0.2s ; 99.9th percentile: 23s
}

\begin{table}[t]
    \centering
    \small
    
\begin{tabularx}{\linewidth}{@{}>{\em}X@{\qquad}SSS[table-format=4]@{}}
    \toprule
    \colorbox{pastelblue!50}{\em all values $\times10^{-3}$} & {Pre-training} & \multicolumn{2}{c}{{Fine-tuning (10\%)}}             \\
    \cmidrule{3-4}
                                                             & {Delay}        & {Delay}                                  & {log MCT} \\
    \midrule

    \mbox{\em{NTT}}                                          &                &                                          &           \\
    \mbox{\smallindent Pre-trained}                          & 0.072          & 0.097                                    & 65        \\
    \mbox{\smallindent From scratch}                         & {-}            & 0.313                                    & 117       \\[1mm]
    \mbox{\em{Baselines}}                                                                                                            \\
    \mbox{\smallindent Last observed}                        & 0.142          & 0.121                                    & 2189      \\
    \mbox{\smallindent EWMA}                                 & 0.259          & 0.211                                    & 1147      \\[1mm]
    \mbox{\em{NTT (Ablated)}}                                                                                                        \\
    \mbox{\smallindent No aggregation}                       & 0.258          & 0.430                                    & 61        \\
    \mbox{\smallindent Fixed aggregation}                    & 0.055          & 0.134                                    & 115       \\[0.75mm]
    \mbox{\smallindent Without packet size}                  & 0.001          & 8.688                                    & 94        \\
    \mbox{\smallindent Without delay}                        & 15.797         & 10.898                                   & 802       \\
    \bottomrule
\end{tabularx}
    \caption{Mean Squared Error for all models and tasks.
        The \emph{pre-trained} NTT outperforms the \emph{from-scratch version}
        and benefits from our design choices (see~\cref{sec:transformer}).
    }
    \label{table:simple-topo}
\end{table}

\paragraph{Case \#1 -- Generalization on the same topology}
We first consider the fine-tuning case 1, where we add unseen cross-traffic on the same topology~(see \cref{fig:ns3}, \cref{table:simple-topo,table:training-time}).

First, we confirm that the \emph{pre-trained} NTT beats all baselines~(\cref{table:simple-topo}).
While this is no breakthrough (the baselines are basic), it suggests that NTT indeed learns sensible values. %

Second, we observe that pre-training is beneficial: on both fine-tuning tasks, the \emph{pre-trained} NTT outperforms the \emph{from scratch} version~(\cref{table:simple-topo});
it generalizes to a new context (\ie unseen cross-traffic) and a new task (\ie MCT prediction).

\begin{table}[h]
    \centering
    \small
    
\begin{tabularx}{\linewidth}{@{}X@{}c@{\;\;}c@{\;\,}c@{}}
    \toprule
                                           & Layers trained & MSE(Delay)                                    & Training time \\

    \midrule
    \mbox{Pre-trained}                            &                & \colorbox{pastelblue!50}{\em $\times10^{-3}$} &               \\
    \mbox{\smallindent \em{Fine-tuning (full)}}   & Decoder only   & 0.033                                         & 8h45          \\
    \mbox{\smallindent \em{Fine-tuning ($10\%$)}} & Decoder only   & 0.037                                         & 3h45          \\
    \mbox{From scratch}                           &                &                                               &               \\
    \mbox{\smallindent \em{Fine-tuning (full)}}   & Full NTT       & 0.036                                         & 26h           \\
    \mbox{\smallindent \em{Fine-tuning ($10\%$)}} & Full NTT       & 0.118                                         & 8h40          \\

    \bottomrule
\end{tabularx}

    \caption{On a simple setting, pre-training saves training resources: fine-tuning data and computing power.}
    \label{table:training-time}
\end{table}

Third, we observe the benefits of hierarchical aggregation and the mix of network and traffic information in the raw data~(\cref{table:simple-topo}).
With \emph{no aggregation}, the model has little history available; we observe that, perhaps surprisingly, this affects the delay predictions but not the MCT ones.
Conversely, with a \emph{fixed aggregation}, the model loses packet-level details but has access to a longer history; this seems sufficient to predict delays but not MCT.
More generally, this initial result suggests that both recent packet-level information and an aggregated history are useful to generalize to a large set of tasks.
Considering the NTT versions \emph{without packet size} and \emph{without delay} information, we observe that neither generalize.
Without packet size, the model overfits the pre-training dataset and performs poorly on predicting delay for fine-tuning.
Without delay information, the model can logically not produce any sensible prediction related to packet delays or MCTs.

Finally, one can argue that the \emph{pre-trained} NTT has an unfair advantage as it trained on about ten times more data than the \emph{from scratch} version.
To put things into perspective, \cref{table:training-time} compares the delay MSE and training time for NTT versions fine-tuned on different datasets.
We observe that fine-tuning on a \emph{full} dataset from scratch yields about the same performance as the on the \emph{10\%} dataset after pre-training.%
\footnote{%
    \cref{table:training-time,table:simple-topo} were obtained with different ``\emph{10\%} fine-tuning datasets''.
    The data allows comparison within each table but not across the two tables.
}
However, fine-tuning on the \emph{full} dataset also requires almost seven times as much training time (26h vs. 3h45).
In practice, collecting fine-tuning data is often expensive; it is thus beneficial to require less.
Finally, fine-tuning from scratch may just not work in more complex settings, as shown next.

\paragraph{Case \#2 -- Generalization on a larger topology}
We now consider the fine-tuning case 2, with several cross-traffic sources on a larger topology~(\cref{fig:ns3}).
In this setting, packets toward different receivers experience different path delays and different levels of congestion from cross-traffic.

\begin{table}[h]
    \centering
    \small
    \begin{tabularx}{\linewidth}{@{}>{\em}X@{}cc@{}}
    \toprule 
    & MSE(Delay)      & Training time                             \\

    \midrule

    \mbox{Pre-trained}  &\colorbox{pastelblue!50}{\em $\times10^{-3}$} & \\
    \mbox{\smallindent \em{Fine-tuning (full)}}  & 0.004      &  10h  \\
    \mbox{\smallindent \em{Fine-tuning ($10\%$)}} & 0.035     &  8h  \\
    \mbox{From scratch}                          &            &   \\
    \mbox{\smallindent \em{Fine-tuning (full)}}   & 5.2       &  20h  \\
    \mbox{\smallindent \em{Fine-tuning ($10\%$)}} & 8.2       &  11h \\
    
    \bottomrule
\end{tabularx}
    \caption{On a larger topology, fine-tuning from scratch no longer works, even if using a large dataset.}
    \label{table:larger-topo}
\end{table}

As evident from \cref{table:larger-topo}, pre-training is essential for NTT to learn basic congestion dynamics first, then generalize later on to the topology's specifics during fine-tuning.%
\footnote{%
    The importance of learning increasingly complex tasks is a problem known as curriculum learning~\cite{narvekar2020Curriculum} and was recently considered in networking~\cite{xiaAutomaticCurriculumGeneration2022}.
}
When fine-tuning \emph{from scratch}, even the \emph{full} dataset is not enough to learn; performance is worse than the baselines~(MSE of 11.2 and 4.0--not shown).
Without addressing information, NTT cannot differentiate between the receivers and thus cannot predict the packet delay accurately~~(MSE of 2.8--not shown).

\section{Conclusion \& Future Research}
\label{sec:future}

Our initial results are promising: they show that NTT effectively learns, that the pre-training knowledge generalizes to new tasks and contexts, and that its specific design benefits overall performance.
Nevertheless, it merely validates that NTT \emph{may work}.
There is a lot more research to assess whether our vision can indeed become a reality.

\paragraph{Does the premise hold?}
We showed some potential of pre-training and fine-tuning with small-scale simulations.
However, real networks are undeniably more complex than this environment.
Real topologies include many paths where many different applications, transport protocols, queuing disciplines, \etc coexist.
There are also many more fine-tuning tasks to consider, \eg flow classification for security or anomaly detection.
Testing our NTT prototype in real, diverse environments and with multiple fine-tuning tasks would provide invaluable insights into the strengths and weaknesses of our architecture and the `learnability' of network dynamics in general.
A next step would be experiments to analyze real-world datasets from Caida~\cite{CAIDAUCSDAnonymized}, M-LAB~\cite{MeasurementLab}, or Crawdad~\cite{Crawdad}.
\researchquestion{How does the NTT hold up with more diverse environments and fine-tuning tasks?\\Which aspects of network dynamics are easy to generalize to, and which are difficult?}

\paragraph{Advancing NTT}
Our prototype architecture~\cite{NTTsoftware} needs enhancements to be helpful in more diverse environments. We see three directions for improvement:
\first packet headers;
\second network telemetry;
and \third sequence aggregation.
Considering packet headers may be essential to learning the behavioral differences of transport protocols or network prioritization of different traffic classes.
However, raw headers are challenging inputs for an ML model, as they may appear in many combinations and contain values that are difficult to learn, like IP addresses~\cite{zhangMimicNetFastPerformance2021}.
Research from the network verification community on header space analysis~\cite{kazemianHeaderSpaceAnalysis2012} may provide valuable insights on header representations and potential first steps in this direction.
In addition, we may collect telemetry data like packet drops or buffer occupancy.
This may help to learn, but not every trace will contain all telemetry, and future research will need to address this potential mismatch.
Finally, we base our prototype aggregation levels on the number of in-flight packets, \ie whether packets in the sequence may share some fate, usually determined by buffer sizes.
The further packets are apart, the less likely they do, and the more we aggregate.
We believe matching individual aggregation levels to typical buffer sizes (\eg flow and switch buffers) may be beneficial. Still, future research needs to put this hypothesis to the test and determine the best sequence sizes and aggregation levels across multiple networks.
\researchquestion{How can we improve the NTT design to learn efficiently from diverse environments? How can we deal with an information mismatch between environments?}

\paragraph{Collaborative pre-training}
Transformers in NLP and CV truly outshone their competition only when pre-trained with massive amounts of data. We envision this could require a previously unseen collaboration across the networking industry.
We see two main challenges:
\first training data volume;
and
\second privacy concerns preventing data sharing.
One can also see these challenges as opportunities:
First, ML models effectively compress data. For example, GPT-3~\cite{brownLanguageModelsAre2020}, one of the largest current Transformer models, consists of 175 Billion parameters or roughly 350 Gigabytes.
However, it contains information from over 45 Terabytes of text data:
Sharing a pre-trained model is much more feasible than sharing all the underlying data, not to mention the savings in training resources.
Second, sharing models instead of data could overcome privacy barriers via federated learning~\cite{kairouzAdvancesOpenProblems2021}:
Organizations could keep their data private and only share pre-trained models, which can be combined into a final collectively pre-trained model.
\researchquestion{Can we leverage pre-training and federated learning to learn from previously unavailable data?}

\paragraph{Continual learning}
A cat remains a cat, but the Internet is an evolving environment.
Protocols, applications, \etc, change over time.
We conjecture that underlying network dynamics change less frequently than specific environments; thus, the same NTT may be used for several updates of the same fine-tuned model.
Nevertheless, even a pre-trained model may become outdated.
It is already difficult to determine when to re-train a specific model~\cite{yanLearningSituRandomized2020}; it might be even more difficult for a model supposed to capture a large range of environments.
\researchquestion{At which point should we consider an NTT outdated? When and with what data should it be re-trained?}

\paragraph{Acknowledgements}
We thank our anonymous reviewers for their helpful comments and feedback.
This work was partially supported by ETH Research Grant ETH-03 19-2.

\message{^^JLASTBODYPAGE \thepage^^J}

\clearpage
\bibliographystyle{ACM-Reference-Format}
\balance
\bibliography{generated_refs}


\begin{thebibliography}{42}


\ifx \showCODEN    \undefined \def \showCODEN     #1{\unskip}     \fi
\ifx \showDOI      \undefined \def \showDOI       #1{#1}\fi
\ifx \showISBNx    \undefined \def \showISBNx     #1{\unskip}     \fi
\ifx \showISBNxiii \undefined \def \showISBNxiii  #1{\unskip}     \fi
\ifx \showISSN     \undefined \def \showISSN      #1{\unskip}     \fi
\ifx \showLCCN     \undefined \def \showLCCN      #1{\unskip}     \fi
\ifx \shownote     \undefined \def \shownote      #1{#1}          \fi
\ifx \showarticletitle \undefined \def \showarticletitle #1{#1}   \fi
\ifx \showURL      \undefined \def \showURL       {\relax}        \fi
\providecommand\bibfield[2]{#2}
\providecommand\bibinfo[2]{#2}
\providecommand\natexlab[1]{#1}
\providecommand\showeprint[2][]{arXiv:#2}

\bibitem[\protect\citeauthoryear{??}{CAI}{2019}]%
        {CAIDAUCSDAnonymized}
 \bibinfo{year}{2019}\natexlab{}.
\newblock \bibinfo{title}{The {{CAIDA UCSD Anonymized Internet Traces}}}.
\newblock   (\bibinfo{year}{2019}).
\newblock
\urldef\tempurl%
\url{http://www.caida.org/data/passive/passive_dataset.xml}
\showURL{%
\tempurl}


\bibitem[\protect\citeauthoryear{??}{Cra}{2022}]%
        {Crawdad}
 \bibinfo{year}{2022}\natexlab{}.
\newblock \bibinfo{title}{Crawdad}.
\newblock   (\bibinfo{year}{2022}).
\newblock
\urldef\tempurl%
\url{https://crawdad.org}
\showURL{%
\tempurl}


\bibitem[\protect\citeauthoryear{??}{Mea}{2022}]%
        {MeasurementLab}
 \bibinfo{year}{2022}\natexlab{}.
\newblock \bibinfo{title}{Measurement {{Lab}}}.
\newblock   (\bibinfo{year}{2022}).
\newblock
\urldef\tempurl%
\url{https://www.measurementlab.net/}
\showURL{%
\tempurl}


\bibitem[\protect\citeauthoryear{Abbasloo, Yen, and Chao}{Abbasloo
  et~al\mbox{.}}{2020}]%
        {abbaslooClassicMeetsModern2020}
\bibfield{author}{\bibinfo{person}{Soheil Abbasloo}, \bibinfo{person}{Chen-Yu
  Yen}, {and} \bibinfo{person}{H.~Jonathan Chao}.}
  \bibinfo{year}{2020}\natexlab{}.
\newblock \showarticletitle{Classic {{Meets Modern}}: {{A Pragmatic
  Learning-Based Congestion Control}} for the {{Internet}}}. In
  \bibinfo{booktitle}{\emph{Proceedings of the {{Annual}} Conference of the
  {{ACM Special Interest Group}} on {{Data Communication}} on the Applications,
  Technologies, Architectures, and Protocols for Computer Communication}}.
  \bibinfo{publisher}{{ACM}}, \bibinfo{address}{{Virtual Event USA}},
  \bibinfo{pages}{632--647}.
\newblock
\showISBNx{978-1-4503-7955-7}
\urldef\tempurl%
\url{https://doi.org/10.1145/3387514.3405892}
\showDOI{\tempurl}


\bibitem[\protect\citeauthoryear{Akhtar, Nam, Govindan, Rao, Chen,
  {Katz-Bassett}, Ribeiro, Zhan, and Zhang}{Akhtar et~al\mbox{.}}{2018}]%
        {akhtarOboeAutotuningVideo2018}
\bibfield{author}{\bibinfo{person}{Zahaib Akhtar}, \bibinfo{person}{Yun~Seong
  Nam}, \bibinfo{person}{Ramesh Govindan}, \bibinfo{person}{Sanjay Rao},
  \bibinfo{person}{Jessica Chen}, \bibinfo{person}{Ethan {Katz-Bassett}},
  \bibinfo{person}{Bruno Ribeiro}, \bibinfo{person}{Jibin Zhan}, {and}
  \bibinfo{person}{Hui Zhang}.} \bibinfo{year}{2018}\natexlab{}.
\newblock \showarticletitle{Oboe: {{Auto-tuning}} Video {{ABR}} Algorithms to
  Network Conditions}. In \bibinfo{booktitle}{\emph{Proceedings of the 2018
  {{Conference}} of the {{ACM Special Interest Group}} on {{Data
  Communication}}}} \emph{(\bibinfo{series}{{{SIGCOMM}} '18})}.
  \bibinfo{publisher}{{Association for Computing Machinery}},
  \bibinfo{address}{{New York, NY, USA}}, \bibinfo{pages}{44--58}.
\newblock
\showISBNx{978-1-4503-5567-4}
\urldef\tempurl%
\url{https://doi.org/10.1145/3230543.3230558}
\showDOI{\tempurl}


\bibitem[\protect\citeauthoryear{Alammar}{Alammar}{2018}]%
        {alammarIllustratedTransformer}
\bibfield{author}{\bibinfo{person}{Jay Alammar}.}
  \bibinfo{year}{2018}\natexlab{}.
\newblock \bibinfo{title}{The {{Illustrated Transformer}}}.
\newblock   (\bibinfo{date}{June} \bibinfo{year}{2018}).
\newblock
\urldef\tempurl%
\url{https://jalammar.github.io/illustrated-transformer/}
\showURL{%
\tempurl}


\bibitem[\protect\citeauthoryear{Bakshy}{Bakshy}{2019}]%
        {bakshyRealworldVideoAdaptation2019}
\bibfield{author}{\bibinfo{person}{Eytan Bakshy}.}
  \bibinfo{year}{2019}\natexlab{}.
\newblock \showarticletitle{Real-World {{Video Adaptation}} with
  {{Reinforcement Learning}}}.
\newblock  (\bibinfo{date}{April} \bibinfo{year}{2019}).
\newblock
\urldef\tempurl%
\url{https://openreview.net/forum?id=SJlCkwN8iV}
\showURL{%
\tempurl}


\bibitem[\protect\citeauthoryear{Bartulovic, Jiang, Balakrishnan, Sekar, and
  Sinopoli}{Bartulovic et~al\mbox{.}}{2017}]%
        {bartulovicBiasesDataDrivenNetworking2017}
\bibfield{author}{\bibinfo{person}{Mihovil Bartulovic},
  \bibinfo{person}{Junchen Jiang}, \bibinfo{person}{Sivaraman Balakrishnan},
  \bibinfo{person}{Vyas Sekar}, {and} \bibinfo{person}{Bruno Sinopoli}.}
  \bibinfo{year}{2017}\natexlab{}.
\newblock \showarticletitle{Biases in {{Data-Driven Networking}}, and {{What}}
  to {{Do About Them}}}. In \bibinfo{booktitle}{\emph{Proceedings of the 16th
  {{ACM Workshop}} on {{Hot Topics}} in {{Networks}}}}
  \emph{(\bibinfo{series}{{{HotNets-XVI}}})}. \bibinfo{publisher}{{Association
  for Computing Machinery}}, \bibinfo{address}{{New York, NY, USA}},
  \bibinfo{pages}{192--198}.
\newblock
\showISBNx{978-1-4503-5569-8}
\urldef\tempurl%
\url{https://doi.org/10.1145/3152434.3152448}
\showDOI{\tempurl}


\bibitem[\protect\citeauthoryear{Beyer, Zhai, and Kolesnikov}{Beyer
  et~al\mbox{.}}{2022}]%
        {beyerBetterPlainViT2022}
\bibfield{author}{\bibinfo{person}{Lucas Beyer}, \bibinfo{person}{Xiaohua
  Zhai}, {and} \bibinfo{person}{Alexander Kolesnikov}.}
  \bibinfo{year}{2022}\natexlab{}.
\newblock \bibinfo{title}{Better Plain {{ViT}} Baselines for {{ImageNet-1k}}}.
\newblock   (\bibinfo{date}{May} \bibinfo{year}{2022}).
\newblock
\urldef\tempurl%
\url{https://doi.org/10.48550/arXiv.2205.01580}
\showDOI{\tempurl}
\showeprint[arxiv]{cs/2205.01580}


\bibitem[\protect\citeauthoryear{Brown, Mann, Ryder, Subbiah, Kaplan, Dhariwal,
  Neelakantan, Shyam, Sastry, Askell, Agarwal, {Herbert-Voss}, Krueger,
  Henighan, Child, Ramesh, Ziegler, Wu, Winter, Hesse, Chen, Sigler, Litwin,
  Gray, Chess, Clark, Berner, McCandlish, Radford, Sutskever, and Amodei}{Brown
  et~al\mbox{.}}{2020}]%
        {brownLanguageModelsAre2020}
\bibfield{author}{\bibinfo{person}{Tom~B. Brown}, \bibinfo{person}{Benjamin
  Mann}, \bibinfo{person}{Nick Ryder}, \bibinfo{person}{Melanie Subbiah},
  \bibinfo{person}{Jared Kaplan}, \bibinfo{person}{Prafulla Dhariwal},
  \bibinfo{person}{Arvind Neelakantan}, \bibinfo{person}{Pranav Shyam},
  \bibinfo{person}{Girish Sastry}, \bibinfo{person}{Amanda Askell},
  \bibinfo{person}{Sandhini Agarwal}, \bibinfo{person}{Ariel {Herbert-Voss}},
  \bibinfo{person}{Gretchen Krueger}, \bibinfo{person}{Tom Henighan},
  \bibinfo{person}{Rewon Child}, \bibinfo{person}{Aditya Ramesh},
  \bibinfo{person}{Daniel~M. Ziegler}, \bibinfo{person}{Jeffrey Wu},
  \bibinfo{person}{Clemens Winter}, \bibinfo{person}{Christopher Hesse},
  \bibinfo{person}{Mark Chen}, \bibinfo{person}{Eric Sigler},
  \bibinfo{person}{Mateusz Litwin}, \bibinfo{person}{Scott Gray},
  \bibinfo{person}{Benjamin Chess}, \bibinfo{person}{Jack Clark},
  \bibinfo{person}{Christopher Berner}, \bibinfo{person}{Sam McCandlish},
  \bibinfo{person}{Alec Radford}, \bibinfo{person}{Ilya Sutskever}, {and}
  \bibinfo{person}{Dario Amodei}.} \bibinfo{year}{2020}\natexlab{}.
\newblock \showarticletitle{Language {{Models}} Are {{Few-Shot Learners}}}.
\newblock \bibinfo{journal}{\emph{arXiv:2005.14165 [cs]}} (\bibinfo{date}{July}
  \bibinfo{year}{2020}).
\newblock
\showeprint[arxiv]{cs/2005.14165}
\urldef\tempurl%
\url{http://arxiv.org/abs/2005.14165}
\showURL{%
\tempurl}


\bibitem[\protect\citeauthoryear{Chen, Lingys, Chen, and Liu}{Chen
  et~al\mbox{.}}{2018}]%
        {chenAuTOScalingDeep2018}
\bibfield{author}{\bibinfo{person}{Li Chen}, \bibinfo{person}{Justinas Lingys},
  \bibinfo{person}{Kai Chen}, {and} \bibinfo{person}{Feng Liu}.}
  \bibinfo{year}{2018}\natexlab{}.
\newblock \showarticletitle{{{AuTO}}: {{Scaling}} Deep Reinforcement Learning
  for Datacenter-Scale Automatic Traffic Optimization}. In
  \bibinfo{booktitle}{\emph{Proceedings of the 2018 {{Conference}} of the {{ACM
  Special Interest Group}} on {{Data Communication}}}}
  \emph{(\bibinfo{series}{{{SIGCOMM}} '18})}. \bibinfo{publisher}{{Association
  for Computing Machinery}}, \bibinfo{address}{{New York, NY, USA}},
  \bibinfo{pages}{191--205}.
\newblock
\showISBNx{978-1-4503-5567-4}
\urldef\tempurl%
\url{https://doi.org/10.1145/3230543.3230551}
\showDOI{\tempurl}


\bibitem[\protect\citeauthoryear{Chen, Gan, Cheng, Liu, and Liu}{Chen
  et~al\mbox{.}}{2020}]%
        {chenDistillingKnowledgeLearned2020}
\bibfield{author}{\bibinfo{person}{Yen-Chun Chen}, \bibinfo{person}{Zhe Gan},
  \bibinfo{person}{Yu Cheng}, \bibinfo{person}{Jingzhou Liu}, {and}
  \bibinfo{person}{Jingjing Liu}.} \bibinfo{year}{2020}\natexlab{}.
\newblock \bibinfo{title}{Distilling {{Knowledge Learned}} in {{BERT}} for
  {{Text Generation}}}.
\newblock   (\bibinfo{date}{July} \bibinfo{year}{2020}).
\newblock
\urldef\tempurl%
\url{https://doi.org/10.48550/arXiv.1911.03829}
\showDOI{\tempurl}
\showeprint[arxiv]{cs/1911.03829}


\bibitem[\protect\citeauthoryear{Devlin, Chang, Lee, and Toutanova}{Devlin
  et~al\mbox{.}}{2019}]%
        {devlinBERTPretrainingDeep2019}
\bibfield{author}{\bibinfo{person}{Jacob Devlin}, \bibinfo{person}{Ming-Wei
  Chang}, \bibinfo{person}{Kenton Lee}, {and} \bibinfo{person}{Kristina
  Toutanova}.} \bibinfo{year}{2019}\natexlab{}.
\newblock \showarticletitle{{{BERT}}: {{Pre-training}} of {{Deep Bidirectional
  Transformers}} for {{Language Understanding}}}.
\newblock \bibinfo{journal}{\emph{arXiv:1810.04805 [cs]}} (\bibinfo{date}{May}
  \bibinfo{year}{2019}).
\newblock
\showeprint[arxiv]{cs/1810.04805}
\urldef\tempurl%
\url{http://arxiv.org/abs/1810.04805}
\showURL{%
\tempurl}


\bibitem[\protect\citeauthoryear{Dosovitskiy, Beyer, Kolesnikov, Weissenborn,
  Zhai, Unterthiner, Dehghani, Minderer, Heigold, Gelly, Uszkoreit, and
  Houlsby}{Dosovitskiy et~al\mbox{.}}{2021}]%
        {dosovitskiyImageWorth16x162021}
\bibfield{author}{\bibinfo{person}{Alexey Dosovitskiy}, \bibinfo{person}{Lucas
  Beyer}, \bibinfo{person}{Alexander Kolesnikov}, \bibinfo{person}{Dirk
  Weissenborn}, \bibinfo{person}{Xiaohua Zhai}, \bibinfo{person}{Thomas
  Unterthiner}, \bibinfo{person}{Mostafa Dehghani}, \bibinfo{person}{Matthias
  Minderer}, \bibinfo{person}{Georg Heigold}, \bibinfo{person}{Sylvain Gelly},
  \bibinfo{person}{Jakob Uszkoreit}, {and} \bibinfo{person}{Neil Houlsby}.}
  \bibinfo{year}{2021}\natexlab{}.
\newblock \showarticletitle{An {{Image}} Is {{Worth}} 16x16 {{Words}}:
  {{Transformers}} for {{Image Recognition}} at {{Scale}}}.
\newblock \bibinfo{journal}{\emph{arXiv:2010.11929 [cs]}} (\bibinfo{date}{June}
  \bibinfo{year}{2021}).
\newblock
\showeprint[arxiv]{cs/2010.11929}
\urldef\tempurl%
\url{http://arxiv.org/abs/2010.11929}
\showURL{%
\tempurl}


\bibitem[\protect\citeauthoryear{Duki{\'c}, Jyothi, Karlas, Owaida, Zhang, and
  Singla}{Duki{\'c} et~al\mbox{.}}{2019}]%
        {dukicAdvanceKnowledgeFlow2019}
\bibfield{author}{\bibinfo{person}{Vojislav Duki{\'c}},
  \bibinfo{person}{Sangeetha~Abdu Jyothi}, \bibinfo{person}{Bojan Karlas},
  \bibinfo{person}{Muhsen Owaida}, \bibinfo{person}{Ce Zhang}, {and}
  \bibinfo{person}{Ankit Singla}.} \bibinfo{year}{2019}\natexlab{}.
\newblock \showarticletitle{Is Advance Knowledge of Flow Sizes a Plausible
  Assumption?}. In \bibinfo{booktitle}{\emph{16th {{USENIX}} Symposium on
  Networked Systems Design and Implementation ({{NSDI}} 19)}}.
  \bibinfo{publisher}{{USENIX Association}}, \bibinfo{address}{{Boston, MA}},
  \bibinfo{pages}{565--580}.
\newblock
\showISBNx{978-1-931971-49-2}
\urldef\tempurl%
\url{https://www.usenix.org/conference/nsdi19/presentation/dukic}
\showURL{%
\tempurl}


\bibitem[\protect\citeauthoryear{Fu, Gupta, Mittal, and Ratnasamy}{Fu
  et~al\mbox{.}}{2021}]%
        {fuUseMLBlackbox2021}
\bibfield{author}{\bibinfo{person}{Silvery Fu}, \bibinfo{person}{Saurabh
  Gupta}, \bibinfo{person}{Radhika Mittal}, {and} \bibinfo{person}{Sylvia
  Ratnasamy}.} \bibinfo{year}{2021}\natexlab{}.
\newblock \showarticletitle{On the {{Use}} of {{ML}} for {{Blackbox System
  Performance Prediction}}}. In \bibinfo{booktitle}{\emph{18th {{USENIX
  Symposium}} on {{Networked Systems Design}} and {{Implementation}} ({{NSDI}}
  21)}}. \bibinfo{pages}{763--784}.
\newblock
\showISBNx{978-1-939133-21-2}
\urldef\tempurl%
\url{https://www.usenix.org/conference/nsdi21/presentation/fu}
\showURL{%
\tempurl}


\bibitem[\protect\citeauthoryear{Han, Wang, Chen, Chen, Guo, Liu, Tang, Xiao,
  Xu, Xu, Yang, Zhang, and Tao}{Han et~al\mbox{.}}{2022}]%
        {hanSurveyVisionTransformer2022}
\bibfield{author}{\bibinfo{person}{Kai Han}, \bibinfo{person}{Yunhe Wang},
  \bibinfo{person}{Hanting Chen}, \bibinfo{person}{Xinghao Chen},
  \bibinfo{person}{Jianyuan Guo}, \bibinfo{person}{Zhenhua Liu},
  \bibinfo{person}{Yehui Tang}, \bibinfo{person}{An Xiao},
  \bibinfo{person}{Chunjing Xu}, \bibinfo{person}{Yixing Xu},
  \bibinfo{person}{Zhaohui Yang}, \bibinfo{person}{Yiman Zhang}, {and}
  \bibinfo{person}{Dacheng Tao}.} \bibinfo{year}{2022}\natexlab{}.
\newblock \showarticletitle{A {{Survey}} on {{Vision Transformer}}}.
\newblock \bibinfo{journal}{\emph{IEEE Transactions on Pattern Analysis and
  Machine Intelligence}} (\bibinfo{year}{2022}), \bibinfo{pages}{1--1}.
\newblock
\showISSN{1939-3539}
\urldef\tempurl%
\url{https://doi.org/10.1109/TPAMI.2022.3152247}
\showDOI{\tempurl}


\bibitem[\protect\citeauthoryear{Huang, Subramanian, Sum, Almubarak, Biderman,
  and Rush}{Huang et~al\mbox{.}}{2022}]%
        {huangAnnotatedTransformer}
\bibfield{author}{\bibinfo{person}{Austin Huang}, \bibinfo{person}{Suraj
  Subramanian}, \bibinfo{person}{Jonathan Sum}, \bibinfo{person}{Khalid
  Almubarak}, \bibinfo{person}{Stella Biderman}, {and} \bibinfo{person}{Sasha
  Rush}.} \bibinfo{year}{2022}\natexlab{}.
\newblock \bibinfo{title}{The {{Annotated Transformer}}}.
\newblock   (\bibinfo{year}{2022}).
\newblock
\urldef\tempurl%
\url{http://nlp.seas.harvard.edu/annotated-transformer/}
\showURL{%
\tempurl}


\bibitem[\protect\citeauthoryear{Jacob}{Jacob}{2022}]%
        {CRediT}
\bibfield{author}{\bibinfo{person}{Romain Jacob}.}
  \bibinfo{year}{2022}\natexlab{}.
\newblock \bibinfo{title}{{{CRediT}} Statement}.
\newblock   (\bibinfo{date}{Oct.} \bibinfo{year}{2022}).
\newblock
\urldef\tempurl%
\url{https://doi.org/10.5281/zenodo.7189024}
\showDOI{\tempurl}


\bibitem[\protect\citeauthoryear{Jay, Rotman, Godfrey, Schapira, and Tamar}{Jay
  et~al\mbox{.}}{2019}]%
        {jayDeepReinforcementLearning2019}
\bibfield{author}{\bibinfo{person}{Nathan Jay}, \bibinfo{person}{Noga Rotman},
  \bibinfo{person}{Brighten Godfrey}, \bibinfo{person}{Michael Schapira}, {and}
  \bibinfo{person}{Aviv Tamar}.} \bibinfo{year}{2019}\natexlab{}.
\newblock \showarticletitle{A {{Deep Reinforcement Learning Perspective}} on
  {{Internet Congestion Control}}}. In \bibinfo{booktitle}{\emph{Proceedings of
  the 36th {{International Conference}} on {{Machine Learning}}}}.
  \bibinfo{publisher}{{PMLR}}, \bibinfo{pages}{3050--3059}.
\newblock
\showISSN{2640-3498}
\urldef\tempurl%
\url{https://proceedings.mlr.press/v97/jay19a.html}
\showURL{%
\tempurl}


\bibitem[\protect\citeauthoryear{Jog, Liu, Franques, Fernando, Abadal,
  Torrellas, and Hassanieh}{Jog et~al\mbox{.}}{2021}]%
        {jogOneProtocolRule2021}
\bibfield{author}{\bibinfo{person}{Suraj Jog}, \bibinfo{person}{Zikun Liu},
  \bibinfo{person}{Antonio Franques}, \bibinfo{person}{Vimuth Fernando},
  \bibinfo{person}{Sergi Abadal}, \bibinfo{person}{Josep Torrellas}, {and}
  \bibinfo{person}{Haitham Hassanieh}.} \bibinfo{year}{2021}\natexlab{}.
\newblock \showarticletitle{One {{Protocol}} to {{Rule Them All}}: {{Wireless}}
  \{\vphantom\}{{Network-on-Chip}}\vphantom\{\} Using {{Deep Reinforcement
  Learning}}}. In \bibinfo{booktitle}{\emph{18th {{USENIX Symposium}} on
  {{Networked Systems Design}} and {{Implementation}} ({{NSDI}} 21)}}.
  \bibinfo{pages}{973--989}.
\newblock
\showISBNx{978-1-939133-21-2}
\urldef\tempurl%
\url{https://www.usenix.org/conference/nsdi21/presentation/jog}
\showURL{%
\tempurl}


\bibitem[\protect\citeauthoryear{Kairouz, McMahan, Avent, Bellet, Bennis,
  Bhagoji, Bonawitz, Charles, Cormode, Cummings, D'Oliveira, Eichner, Rouayheb,
  Evans, Gardner, Garrett, Gasc{\'o}n, Ghazi, Gibbons, Gruteser, Harchaoui, He,
  He, Huo, Hutchinson, Hsu, Jaggi, Javidi, Joshi, Khodak, Konecn{\'y},
  Korolova, Koushanfar, Koyejo, Lepoint, Liu, Mittal, Mohri, Nock,
  {\"O}zg{\"u}r, Pagh, Qi, Ramage, Raskar, Raykova, Song, Song, Stich, Sun,
  Suresh, Tram{\`e}r, Vepakomma, Wang, Xiong, Xu, Yang, Yu, Yu, and
  Zhao}{Kairouz et~al\mbox{.}}{2021}]%
        {kairouzAdvancesOpenProblems2021}
\bibfield{author}{\bibinfo{person}{Peter Kairouz}, \bibinfo{person}{H.~Brendan
  McMahan}, \bibinfo{person}{Brendan Avent}, \bibinfo{person}{Aur{\'e}lien
  Bellet}, \bibinfo{person}{Mehdi Bennis}, \bibinfo{person}{Arjun~Nitin
  Bhagoji}, \bibinfo{person}{Kallista Bonawitz}, \bibinfo{person}{Zachary
  Charles}, \bibinfo{person}{Graham Cormode}, \bibinfo{person}{Rachel
  Cummings}, \bibinfo{person}{Rafael G.~L. D'Oliveira}, \bibinfo{person}{Hubert
  Eichner}, \bibinfo{person}{Salim~El Rouayheb}, \bibinfo{person}{David Evans},
  \bibinfo{person}{Josh Gardner}, \bibinfo{person}{Zachary Garrett},
  \bibinfo{person}{Adri{\`a} Gasc{\'o}n}, \bibinfo{person}{Badih Ghazi},
  \bibinfo{person}{Phillip~B. Gibbons}, \bibinfo{person}{Marco Gruteser},
  \bibinfo{person}{Zaid Harchaoui}, \bibinfo{person}{Chaoyang He},
  \bibinfo{person}{Lie He}, \bibinfo{person}{Zhouyuan Huo},
  \bibinfo{person}{Ben Hutchinson}, \bibinfo{person}{Justin Hsu},
  \bibinfo{person}{Martin Jaggi}, \bibinfo{person}{Tara Javidi},
  \bibinfo{person}{Gauri Joshi}, \bibinfo{person}{Mikhail Khodak},
  \bibinfo{person}{Jakub Konecn{\'y}}, \bibinfo{person}{Aleksandra Korolova},
  \bibinfo{person}{Farinaz Koushanfar}, \bibinfo{person}{Sanmi Koyejo},
  \bibinfo{person}{Tancr{\`e}de Lepoint}, \bibinfo{person}{Yang Liu},
  \bibinfo{person}{Prateek Mittal}, \bibinfo{person}{Mehryar Mohri},
  \bibinfo{person}{Richard Nock}, \bibinfo{person}{Ayfer {\"O}zg{\"u}r},
  \bibinfo{person}{Rasmus Pagh}, \bibinfo{person}{Hang Qi},
  \bibinfo{person}{Daniel Ramage}, \bibinfo{person}{Ramesh Raskar},
  \bibinfo{person}{Mariana Raykova}, \bibinfo{person}{Dawn Song},
  \bibinfo{person}{Weikang Song}, \bibinfo{person}{Sebastian~U. Stich},
  \bibinfo{person}{Ziteng Sun}, \bibinfo{person}{Ananda~Theertha Suresh},
  \bibinfo{person}{Florian Tram{\`e}r}, \bibinfo{person}{Praneeth Vepakomma},
  \bibinfo{person}{Jianyu Wang}, \bibinfo{person}{Li Xiong},
  \bibinfo{person}{Zheng Xu}, \bibinfo{person}{Qiang Yang},
  \bibinfo{person}{Felix~X. Yu}, \bibinfo{person}{Han Yu}, {and}
  \bibinfo{person}{Sen Zhao}.} \bibinfo{year}{2021}\natexlab{}.
\newblock \showarticletitle{Advances and {{Open Problems}} in {{Federated
  Learning}}}.
\newblock \bibinfo{journal}{\emph{Foundations and Trends\textregistered{} in
  Machine Learning}} \bibinfo{volume}{14}, \bibinfo{number}{1\textendash 2}
  (\bibinfo{date}{June} \bibinfo{year}{2021}), \bibinfo{pages}{1--210}.
\newblock
\showISSN{1935-8237, 1935-8245}
\urldef\tempurl%
\url{https://doi.org/10.1561/2200000083}
\showDOI{\tempurl}


\bibitem[\protect\citeauthoryear{Kazemian, Varghese, and McKeown}{Kazemian
  et~al\mbox{.}}{2012}]%
        {kazemianHeaderSpaceAnalysis2012}
\bibfield{author}{\bibinfo{person}{Peyman Kazemian}, \bibinfo{person}{George
  Varghese}, {and} \bibinfo{person}{Nick McKeown}.}
  \bibinfo{year}{2012}\natexlab{}.
\newblock \showarticletitle{Header {{Space Analysis}}: {{Static Checking}} for
  {{Networks}}}. In \bibinfo{booktitle}{\emph{9th {{USENIX Symposium}} on
  {{Networked Systems Design}} and {{Implementation}} ({{NSDI}} 12)}}.
  \bibinfo{pages}{113--126}.
\newblock


\bibitem[\protect\citeauthoryear{Khan, Naseer, Hayat, Zamir, Khan, and
  Shah}{Khan et~al\mbox{.}}{2021}]%
        {khanTransformersVisionSurvey2021}
\bibfield{author}{\bibinfo{person}{Salman Khan}, \bibinfo{person}{Muzammal
  Naseer}, \bibinfo{person}{Munawar Hayat}, \bibinfo{person}{Syed~Waqas Zamir},
  \bibinfo{person}{Fahad~Shahbaz Khan}, {and} \bibinfo{person}{Mubarak Shah}.}
  \bibinfo{year}{2021}\natexlab{}.
\newblock \showarticletitle{Transformers in {{Vision}}: {{A Survey}}}.
\newblock \bibinfo{journal}{\emph{Comput. Surveys}} (\bibinfo{date}{Dec.}
  \bibinfo{year}{2021}).
\newblock
\showISSN{0360-0300}
\urldef\tempurl%
\url{https://doi.org/10.1145/3505244}
\showDOI{\tempurl}


\bibitem[\protect\citeauthoryear{Mao, Netravali, and Alizadeh}{Mao
  et~al\mbox{.}}{2017}]%
        {maoNeuralAdaptiveVideo2017}
\bibfield{author}{\bibinfo{person}{Hongzi Mao}, \bibinfo{person}{Ravi
  Netravali}, {and} \bibinfo{person}{Mohammad Alizadeh}.}
  \bibinfo{year}{2017}\natexlab{}.
\newblock \showarticletitle{Neural {{Adaptive Video Streaming}} with
  {{Pensieve}}}. In \bibinfo{booktitle}{\emph{Proceedings of the {{Conference}}
  of the {{ACM Special Interest Group}} on {{Data Communication}}}}
  \emph{(\bibinfo{series}{{{SIGCOMM}} '17})}. \bibinfo{publisher}{{Association
  for Computing Machinery}}, \bibinfo{address}{{Los Angeles, CA, USA}},
  \bibinfo{pages}{197--210}.
\newblock
\showISBNx{978-1-4503-4653-5}
\urldef\tempurl%
\url{https://doi.org/10.1145/3098822.3098843}
\showDOI{\tempurl}


\bibitem[\protect\citeauthoryear{Montazeri, Li, Alizadeh, and
  Ousterhout}{Montazeri et~al\mbox{.}}{2018}]%
        {montazeriHomaReceiverdrivenLowlatency2018}
\bibfield{author}{\bibinfo{person}{Behnam Montazeri}, \bibinfo{person}{Yilong
  Li}, \bibinfo{person}{Mohammad Alizadeh}, {and} \bibinfo{person}{John
  Ousterhout}.} \bibinfo{year}{2018}\natexlab{}.
\newblock \showarticletitle{Homa: {{A}} Receiver-Driven Low-Latency Transport
  Protocol Using Network Priorities}. In \bibinfo{booktitle}{\emph{Proceedings
  of the 2018 {{Conference}} of the {{ACM Special Interest Group}} on {{Data
  Communication}}}} \emph{(\bibinfo{series}{{{SIGCOMM}} '18})}.
  \bibinfo{publisher}{{Association for Computing Machinery}},
  \bibinfo{address}{{Budapest, Hungary}}, \bibinfo{pages}{221--235}.
\newblock
\showISBNx{978-1-4503-5567-4}
\urldef\tempurl%
\url{https://doi.org/10.1145/3230543.3230564}
\showDOI{\tempurl}


\bibitem[\protect\citeauthoryear{Narvekar, Peng, Leonetti, Sinapov, Taylor, and
  Stone}{Narvekar et~al\mbox{.}}{2020}]%
        {narvekar2020Curriculum}
\bibfield{author}{\bibinfo{person}{Sanmit Narvekar}, \bibinfo{person}{Bei
  Peng}, \bibinfo{person}{Matteo Leonetti}, \bibinfo{person}{Jivko Sinapov},
  \bibinfo{person}{Matthew~E. Taylor}, {and} \bibinfo{person}{Peter Stone}.}
  \bibinfo{year}{2020}\natexlab{}.
\newblock \bibinfo{title}{Curriculum {{Learning}} for {{Reinforcement Learning
  Domains}}: {{A Framework}} and {{Survey}}}.
\newblock   (\bibinfo{date}{Sept.} \bibinfo{year}{2020}).
\newblock
\urldef\tempurl%
\url{https://doi.org/10.48550/arXiv.2003.04960}
\showDOI{\tempurl}
\showeprint[arxiv]{cs, stat/2003.04960}


\bibitem[\protect\citeauthoryear{Nie, Zhao, Li, Chen, Sui, Zhang, Ye, and
  Pei}{Nie et~al\mbox{.}}{2019}]%
        {nieDynamicTCPInitial2019}
\bibfield{author}{\bibinfo{person}{X. Nie}, \bibinfo{person}{Y. Zhao},
  \bibinfo{person}{Z. Li}, \bibinfo{person}{G. Chen}, \bibinfo{person}{K. Sui},
  \bibinfo{person}{J. Zhang}, \bibinfo{person}{Z. Ye}, {and}
  \bibinfo{person}{D. Pei}.} \bibinfo{year}{2019}\natexlab{}.
\newblock \showarticletitle{Dynamic {{TCP Initial Windows}} and {{Congestion
  Control Schemes Through Reinforcement Learning}}}.
\newblock \bibinfo{journal}{\emph{IEEE Journal on Selected Areas in
  Communications}} \bibinfo{volume}{37}, \bibinfo{number}{6}
  (\bibinfo{date}{June} \bibinfo{year}{2019}), \bibinfo{pages}{1231--1247}.
\newblock
\showISSN{1558-0008}
\urldef\tempurl%
\url{https://doi.org/10.1109/JSAC.2019.2904350}
\showDOI{\tempurl}


\bibitem[\protect\citeauthoryear{Poupart, Chen, Jaini, Fung, Susanto, {Yanhui
  Geng}, {Li Chen}, Chen, and {Hao Jin}}{Poupart et~al\mbox{.}}{2016}]%
        {poupartOnlineFlowSize2016}
\bibfield{author}{\bibinfo{person}{P. Poupart}, \bibinfo{person}{Z. Chen},
  \bibinfo{person}{P. Jaini}, \bibinfo{person}{F. Fung}, \bibinfo{person}{H.
  Susanto}, \bibinfo{person}{{Yanhui Geng}}, \bibinfo{person}{{Li Chen}},
  \bibinfo{person}{K. Chen}, {and} \bibinfo{person}{{Hao Jin}}.}
  \bibinfo{year}{2016}\natexlab{}.
\newblock \showarticletitle{Online Flow Size Prediction for Improved Network
  Routing}. In \bibinfo{booktitle}{\emph{2016 {{IEEE}} 24th {{International
  Conference}} on {{Network Protocols}} ({{ICNP}})}}. \bibinfo{pages}{1--6}.
\newblock
\urldef\tempurl%
\url{https://doi.org/10.1109/ICNP.2016.7785324}
\showDOI{\tempurl}


\bibitem[\protect\citeauthoryear{Ray and Dietm{\"u}ller}{Ray and
  Dietm{\"u}ller}{2022}]%
        {NTTsoftware}
\bibfield{author}{\bibinfo{person}{Siddhant Ray} {and}
  \bibinfo{person}{Alexander Dietm{\"u}ller}.} \bibinfo{year}{2022}\natexlab{}.
\newblock \bibinfo{title}{Network Traffic Transformer}.
\newblock \bibinfo{howpublished}{Zenodo}.   (\bibinfo{date}{Oct.}
  \bibinfo{year}{2022}).
\newblock
\urldef\tempurl%
\url{https://doi.org/10.5281/zenodo.7186893}
\showDOI{\tempurl}


\bibitem[\protect\citeauthoryear{Riley and Henderson}{Riley and
  Henderson}{2010}]%
        {rileyNs3NetworkSimulator2010}
\bibfield{author}{\bibinfo{person}{George~F. Riley} {and}
  \bibinfo{person}{Thomas~R. Henderson}.} \bibinfo{year}{2010}\natexlab{}.
\newblock \showarticletitle{The Ns-3 {{Network Simulator}}}.
\newblock In \bibinfo{booktitle}{\emph{Modeling and {{Tools}} for {{Network
  Simulation}}}}, \bibfield{editor}{\bibinfo{person}{Klaus Wehrle},
  \bibinfo{person}{Mesut G{\"u}ne{\c s}}, {and} \bibinfo{person}{James Gross}}
  (Eds.). \bibinfo{publisher}{{Springer}}, \bibinfo{address}{{Berlin,
  Heidelberg}}, \bibinfo{pages}{15--34}.
\newblock
\showISBNx{978-3-642-12331-3}
\urldef\tempurl%
\url{https://doi.org/10.1007/978-3-642-12331-3_2}
\showDOI{\tempurl}


\bibitem[\protect\citeauthoryear{Sivaraman, Winstein, Thaker, and
  Balakrishnan}{Sivaraman et~al\mbox{.}}{2014}]%
        {sivaramanExperimentalStudyLearnability2014}
\bibfield{author}{\bibinfo{person}{Anirudh Sivaraman}, \bibinfo{person}{Keith
  Winstein}, \bibinfo{person}{Pratiksha Thaker}, {and} \bibinfo{person}{Hari
  Balakrishnan}.} \bibinfo{year}{2014}\natexlab{}.
\newblock \showarticletitle{An {{Experimental Study}} of the {{Learnability}}
  of {{Congestion Control}}}. In \bibinfo{booktitle}{\emph{Proceedings of the
  2014 {{ACM Conference}} on {{SIGCOMM}}}} \emph{(\bibinfo{series}{{{SIGCOMM}}
  '14})}. \bibinfo{publisher}{{ACM}}, \bibinfo{address}{{New York, NY, USA}},
  \bibinfo{pages}{479--490}.
\newblock
\showISBNx{978-1-4503-2836-4}
\urldef\tempurl%
\url{https://doi.org/10.1145/2619239.2626324}
\showDOI{\tempurl}


\bibitem[\protect\citeauthoryear{Storks, Gao, and Chai}{Storks
  et~al\mbox{.}}{2020}]%
        {storksRecentAdvancesNatural2020}
\bibfield{author}{\bibinfo{person}{Shane Storks}, \bibinfo{person}{Qiaozi Gao},
  {and} \bibinfo{person}{Joyce~Y. Chai}.} \bibinfo{year}{2020}\natexlab{}.
\newblock \bibinfo{title}{Recent {{Advances}} in {{Natural Language
  Inference}}: {{A Survey}} of {{Benchmarks}}, {{Resources}}, and
  {{Approaches}}}.
\newblock   (\bibinfo{date}{Feb.} \bibinfo{year}{2020}).
\newblock
\urldef\tempurl%
\url{https://doi.org/10.48550/arXiv.1904.01172}
\showDOI{\tempurl}
\showeprint[arxiv]{cs/1904.01172}


\bibitem[\protect\citeauthoryear{Valadarsky, Schapira, Shahaf, and
  Tamar}{Valadarsky et~al\mbox{.}}{2017}]%
        {valadarskyLearningRoute2017}
\bibfield{author}{\bibinfo{person}{Asaf Valadarsky}, \bibinfo{person}{Michael
  Schapira}, \bibinfo{person}{Dafna Shahaf}, {and} \bibinfo{person}{Aviv
  Tamar}.} \bibinfo{year}{2017}\natexlab{}.
\newblock \showarticletitle{Learning to {{Route}}}. In
  \bibinfo{booktitle}{\emph{Proceedings of the 16th {{ACM Workshop}} on {{Hot
  Topics}} in {{Networks}}}} \emph{(\bibinfo{series}{{{HotNets-XVI}}})}.
  \bibinfo{publisher}{{ACM}}, \bibinfo{address}{{New York, NY, USA}},
  \bibinfo{pages}{185--191}.
\newblock
\showISBNx{978-1-4503-5569-8}
\urldef\tempurl%
\url{https://doi.org/10.1145/3152434.3152441}
\showDOI{\tempurl}


\bibitem[\protect\citeauthoryear{Vaswani, Shazeer, Parmar, Uszkoreit, Jones,
  Gomez, Kaiser, and Polosukhin}{Vaswani et~al\mbox{.}}{2017}]%
        {vaswaniAttentionAllYou2017}
\bibfield{author}{\bibinfo{person}{Ashish Vaswani}, \bibinfo{person}{Noam
  Shazeer}, \bibinfo{person}{Niki Parmar}, \bibinfo{person}{Jakob Uszkoreit},
  \bibinfo{person}{Llion Jones}, \bibinfo{person}{Aidan~N Gomez},
  \bibinfo{person}{{\L}ukasz Kaiser}, {and} \bibinfo{person}{Illia
  Polosukhin}.} \bibinfo{year}{2017}\natexlab{}.
\newblock \showarticletitle{Attention Is All You Need}. In
  \bibinfo{booktitle}{\emph{Advances in Neural Information Processing
  Systems}}, \bibfield{editor}{\bibinfo{person}{I.~Guyon},
  \bibinfo{person}{U.~V. Luxburg}, \bibinfo{person}{S.~Bengio},
  \bibinfo{person}{H.~Wallach}, \bibinfo{person}{R.~Fergus},
  \bibinfo{person}{S.~Vishwanathan}, {and} \bibinfo{person}{R.~Garnett}}
  (Eds.), Vol.~\bibinfo{volume}{30}. \bibinfo{publisher}{{Curran Associates,
  Inc.}}
\newblock
\urldef\tempurl%
\url{https://proceedings.neurips.cc/paper/2017/file/3f5ee243547dee91fbd053c1c4a845aa-Paper.pdf}
\showURL{%
\tempurl}


\bibitem[\protect\citeauthoryear{Winstein and Balakrishnan}{Winstein and
  Balakrishnan}{2013}]%
        {winsteinTCPExMachina2013}
\bibfield{author}{\bibinfo{person}{Keith Winstein} {and} \bibinfo{person}{Hari
  Balakrishnan}.} \bibinfo{year}{2013}\natexlab{}.
\newblock \showarticletitle{{{TCP Ex Machina}}: {{Computer-generated Congestion
  Control}}}. In \bibinfo{booktitle}{\emph{Proceedings of the {{ACM SIGCOMM}}
  2013 {{Conference}} on {{SIGCOMM}}}} \emph{(\bibinfo{series}{{{SIGCOMM}}
  '13})}. \bibinfo{publisher}{{ACM}}, \bibinfo{address}{{New York, NY, USA}},
  \bibinfo{pages}{123--134}.
\newblock
\showISBNx{978-1-4503-2056-6}
\urldef\tempurl%
\url{https://doi.org/10.1145/2486001.2486020}
\showDOI{\tempurl}


\bibitem[\protect\citeauthoryear{Xia, Zhou, Yan, and Jiang}{Xia
  et~al\mbox{.}}{2022}]%
        {xiaAutomaticCurriculumGeneration2022}
\bibfield{author}{\bibinfo{person}{Zhengxu Xia}, \bibinfo{person}{Yajie Zhou},
  \bibinfo{person}{Francis~Y. Yan}, {and} \bibinfo{person}{Junchen Jiang}.}
  \bibinfo{year}{2022}\natexlab{}.
\newblock \bibinfo{title}{Automatic {{Curriculum Generation}} for {{Learning
  Adaptation}} in {{Networking}}}.
\newblock   (\bibinfo{date}{Sept.} \bibinfo{year}{2022}).
\newblock
\urldef\tempurl%
\url{https://doi.org/10.48550/arXiv.2202.05940}
\showDOI{\tempurl}
\showeprint[arxiv]{cs/2202.05940}


\bibitem[\protect\citeauthoryear{Yan, Ayers, Zhu, Fouladi, Hong, Zhang, Levis,
  and Winstein}{Yan et~al\mbox{.}}{2020}]%
        {yanLearningSituRandomized2020}
\bibfield{author}{\bibinfo{person}{Francis~Y. Yan}, \bibinfo{person}{Hudson
  Ayers}, \bibinfo{person}{Chenzhi Zhu}, \bibinfo{person}{Sadjad Fouladi},
  \bibinfo{person}{James Hong}, \bibinfo{person}{Keyi Zhang},
  \bibinfo{person}{Philip Levis}, {and} \bibinfo{person}{Keith Winstein}.}
  \bibinfo{year}{2020}\natexlab{}.
\newblock \showarticletitle{Learning in Situ: {{A}} Randomized Experiment in
  Video Streaming}. In \bibinfo{booktitle}{\emph{17th {{USENIX Symposium}} on
  {{Networked Systems Design}} and {{Implementation}} ({{NSDI}} 20)}}.
  \bibinfo{pages}{495--511}.
\newblock
\showISBNx{978-1-939133-13-7}
\urldef\tempurl%
\url{https://www.usenix.org/conference/nsdi20/presentation/yan}
\showURL{%
\tempurl}


\bibitem[\protect\citeauthoryear{Yan, Ma, Hill, Raghavan, Wahby, Levis, and
  Winstein}{Yan et~al\mbox{.}}{2018}]%
        {yanPantheonTrainingGround2018}
\bibfield{author}{\bibinfo{person}{Francis~Y. Yan}, \bibinfo{person}{Jestin
  Ma}, \bibinfo{person}{Greg~D. Hill}, \bibinfo{person}{Deepti Raghavan},
  \bibinfo{person}{Riad~S. Wahby}, \bibinfo{person}{Philip Levis}, {and}
  \bibinfo{person}{Keith Winstein}.} \bibinfo{year}{2018}\natexlab{}.
\newblock \showarticletitle{Pantheon: {{The}} Training Ground for {{Internet}}
  Congestion-Control Research}. In \bibinfo{booktitle}{\emph{2018 {{USENIX
  Annual Technical Conference}} ({{USENIX ATC}} 18)}}.
  \bibinfo{pages}{731--743}.
\newblock
\showISBNx{978-1-931971-44-7}
\urldef\tempurl%
\url{https://www.usenix.org/conference/atc18/presentation/yan-francis}
\showURL{%
\tempurl}


\bibitem[\protect\citeauthoryear{Yu, Wang, and Liew}{Yu et~al\mbox{.}}{2019}]%
        {yuDeepReinforcementLearningMultiple2019}
\bibfield{author}{\bibinfo{person}{Yiding Yu}, \bibinfo{person}{Taotao Wang},
  {and} \bibinfo{person}{Soung~Chang Liew}.} \bibinfo{year}{2019}\natexlab{}.
\newblock \showarticletitle{Deep-{{Reinforcement Learning Multiple Access}} for
  {{Heterogeneous Wireless Networks}}}.
\newblock \bibinfo{journal}{\emph{IEEE Journal on Selected Areas in
  Communications}} \bibinfo{volume}{37}, \bibinfo{number}{6}
  (\bibinfo{date}{June} \bibinfo{year}{2019}), \bibinfo{pages}{1277--1290}.
\newblock
\showISSN{1558-0008}
\urldef\tempurl%
\url{https://doi.org/10.1109/JSAC.2019.2904329}
\showDOI{\tempurl}


\bibitem[\protect\citeauthoryear{Zaib, Tran, Sagar, Mahmood, Zhang, and
  Sheng}{Zaib et~al\mbox{.}}{2021}]%
        {zaibBERTCoQACBERTbasedConversational2021}
\bibfield{author}{\bibinfo{person}{Munazza Zaib}, \bibinfo{person}{Dai~Hoang
  Tran}, \bibinfo{person}{Subhash Sagar}, \bibinfo{person}{Adnan Mahmood},
  \bibinfo{person}{Wei~E. Zhang}, {and} \bibinfo{person}{Quan~Z. Sheng}.}
  \bibinfo{year}{2021}\natexlab{}.
\newblock \bibinfo{title}{{{BERT-CoQAC}}: {{BERT-based Conversational Question
  Answering}} in {{Context}}}.
\newblock   (\bibinfo{date}{April} \bibinfo{year}{2021}).
\newblock
\urldef\tempurl%
\url{https://doi.org/10.48550/arXiv.2104.11394}
\showDOI{\tempurl}
\showeprint[arxiv]{cs/2104.11394}


\bibitem[\protect\citeauthoryear{Zhang, Ng, Kazer, Yan, Sedoc, and Liu}{Zhang
  et~al\mbox{.}}{2021}]%
        {zhangMimicNetFastPerformance2021}
\bibfield{author}{\bibinfo{person}{Qizhen Zhang}, \bibinfo{person}{Kelvin K.~W.
  Ng}, \bibinfo{person}{Charles Kazer}, \bibinfo{person}{Shen Yan},
  \bibinfo{person}{Jo{\~a}o Sedoc}, {and} \bibinfo{person}{Vincent Liu}.}
  \bibinfo{year}{2021}\natexlab{}.
\newblock \showarticletitle{{{MimicNet}}: {{Fast}} Performance Estimates for
  Data Center Networks with Machine Learning}. In
  \bibinfo{booktitle}{\emph{Proceedings of the 2021 {{ACM SIGCOMM}} 2021
  {{Conference}}}} \emph{(\bibinfo{series}{{{SIGCOMM}} '21})}.
  \bibinfo{publisher}{{Association for Computing Machinery}},
  \bibinfo{address}{{New York, NY, USA}}, \bibinfo{pages}{287--304}.
\newblock
\showISBNx{978-1-4503-8383-7}
\urldef\tempurl%
\url{https://doi.org/10.1145/3452296.3472926}
\showDOI{\tempurl}


\end{thebibliography}

\message{^^JLASTREFERENCESPAGE \thepage^^J}

\ifbool{includeappendix}{%
  
}{}

\end{document}